\documentstyle [11pt]{article}
\input amssym.def
\input amssym.tex
\newtheorem{th}{Theorem}[section] 

\newtheorem{prop}[th]{Proposition} 
\newtheorem{lem}[th]{Lemma}

\newcommand{\sect}[1]{\setcounter{equation}{0}\section{#1}}
 
\def\N{\mbox{I\hspace{-.15em}N}}

\def\Z{\mbox{Z\hspace{-.3em}Z}}

\def\R{{\rm I\hspace{-.15em}R}}

\def\Q{\mbox{l\hspace{-.47em}Q}}

\def\To{ {\bf S}^1 }

\def\endproof{\hfill$\square$}

\begin{document}
\title{Transport properties of kicked and quasi-periodic Hamiltonians}   
\author{S. De Bi\`evre\thanks{E-mail: debievre@gat.univ-lille1.fr}\\ 
UFR de Math\'ematiques et URA GAT\\
Universit\'e des Sciences et Technologies de Lille\\
59655 Villeneuve d'Ascq Cedex France \and
G. Forni\thanks{E-mail: gforni@math.princeton.edu}\\
Department of Mathematics\\
Princeton University\\ Princeton 08544 NJ USA }

\date{April 23 1997}

\maketitle
\centerline{to appear in the Journal of Statistical Physics}

\vskip 0.5cm

\begin{abstract} 
{\em We study transport properties of Schr\"odinger operators
depending on one or more parameters.  Examples include 
the kicked rotor and operators with quasi-periodic potentials.
We show that  the mean growth exponent of the
kinetic energy  in the kicked 
rotor and of the mean square displacement in quasi-periodic potentials is
generically equal to 2: this means that the motion remains ballistic, at 
least in a weak sense, even away from the resonances of the models. 
Stronger results are 
obtained for a class of tight-binding Hamiltonians with an
 electric field $E(t)= E_0 + E_1\cos\omega t$. For
$$
H=\sum a_{n-k}(\mid n-k><n\mid + \mid n>< n-k\mid) + E(t)\mid n><n\mid
$$
with $a_n\sim\mid n\mid^{-\nu}\ (\nu>3/2)$ we show
that the mean square displacement satisfies $\overline{<\psi_t, N^2\psi_t>}\geq
C_\epsilon t^{2/(\nu+1/2)-\epsilon}$ for suitable choices of $\omega, E_0$ and 
$E_1$.
We relate this behaviour to the spectral properties of the Floquet operator of
the problem.

} \end{abstract}

\sect{Introduction and statement of the results}
The asymptotic behaviour of time dependent quantities in the study of 
Schr\"odinger equations with quasi-periodic and random potentials, or in the
study of periodically kicked or pulsed Hamiltonian systems has attracted much
attention in the last decade.  For a review we refer to [Ho] where many
references beyond the ones cited below can be found. We present in section 2 two 
simple abstract results stating roughly that ``close to resonance" the spectrum of the
Hamiltonian or the Floquet operator is continuous and that the motion has
the same asymptotic characteristics as ``in resonance", at least in some weak
sense.  In section 3, we first apply these results to the kicked rotor.  We give
a simplified proof of the Casati-Guarneri result stating that generically the
spectrum of the Floquet operator of the kicked rotor is continuous.  We show in
addition that
the mean growth exponent of the time-averaged kinetic energy, defined below,
generically equals 2 (Theorem 3.1). This can be paraphrased (somewhat exaggeratedly) by
saying that the time-averaged  kinetic energy  generically behaves
ballistically in the kicked rotor.  We then show how to prove similar results
for  quasi-periodic
Schr\"odinger operators.  In section 4, we study in detail the
asymptotic behaviour of the kicked linear rotor.  We exhibit various
phenomena relating the asymptotic behaviour of the time-averaged kinetic
energy and the nature of the  spectrum of the Floquet operator, strenghtening
known results.   In section 5 we then show how these results apply to the
problem of the motion of a charged particle in a one-band tight-binding model
subjected to a time-dependent electric field (Theorem 5.1).  We now first describe our
results and the motivations in some more detail.  

Consider a periodically driven classical or quantal Hamiltonian system with 
Hamiltonian 
$$
H(t) = H_0 + V(t),\quad V(t+T) = V(t).
$$
In classical mechanics the evolution of the system is determined by the Floquet
transformation $\Phi_V$, obtained by integrating the Hamiltonian equations of
motion over one period $T$.  It is a canonical transformation of phase space
$M$.  Similarly, in quantum mechanics, integration of the Schr\"odinger equation
over one period $T$ yields the Floquet operator $U_V$, which is a unitary
operator on the Hilbert space of states ${\cal H}$.  Typically, the
unperturbed classical Hamiltonian is chosen to be completely integrable with its
motion restricted to invariant tori, so that in particular all its
trajectories are bounded.  The corresponding quantum Hamiltonian is assumed to
have pure point spectrum so that the same is true for $U_0=\exp-{i\over\hbar}
TH_0$.  What happens when $V$ is turned on? 

A natural first question to ask is whether the unperturbed energy $H_0$, which
is a constant of the motion when $V=0$, remains bounded when $V\not=0$. More
precisely, the question in classical mechanics is whether $$
\sup_{m\in\Z}\mid H_0\circ\Phi_V^m(x,p)\mid<\infty,\ (x,p)\in M,
$$
and in quantum mechanics whether   
$$
\sup_{m\in\Z}\mid <\psi,U_V^{-m}H_0U_V^m\psi>\mid<\infty, \psi\in
{\cal D}(H_0), $$
where ${\cal D}(H_0)$ denotes the domain of $H_0$, assumed stable under
$U_V$.   Without giving a precise definition, let us say the system is 
 dynamically stable
in this case, and dynamically unstable otherwise. 

In quantum mechanics one can ask a different, but related stability question
[Be][Ho]: is the spectrum of the Floquet operator $U_V$ still pure point? We
will say the system is spectrally stable if this is the case. It is a
well known consequence of the RAGE theorem [CFKS] that dynamical stability
implies spectral stability, the opposite implication not being true [DJLS].  

We will be interested in dynamically unstable systems.  Once the unperturbed
energy does not remain bounded under the full, perturbed evolution, one can
ask about its asymptotic behaviour: how do  $\mid H_0\circ\Phi_V^m(x,p)\mid$
and $\mid <\psi,U_V^{-m}H_0U_V^m\psi>\mid$  behave as $m$ goes to infinity?
One is in particular interested in finding out whether these quantities can
have algebraic growth and, in the quantum mechanical case, to relate the
growth exponent to the spectral properties of $U_V$. 

Before reviewing the known results, let us quickly recall the various notions
of growth exponent one might consider. Let $h_m$ be a sequence of positive
numbers. One says the sequence displays algebraic growth with exponent
$\alpha(h)\geq 0$ if there exist constants $c$ and $C$ so that, for all
$m\in\N^*$ \begin{equation} 
c\ m^{\alpha(h)} \leq h_m \leq C\ m^{\alpha(h)}.
\end{equation}
It turns out this notion is too strong to be of use: the quantities of interest
tend to fluctuate greatly as we will show in detail for the kicked linear rotor
in section 4.  To obtain a fluctuation independent quantity, it has been
suggested [Gu] one should consider a ``mean growth exponent" $\alpha_0 (h)$
defined as \begin{eqnarray}
\alpha_0(h) &=&\inf \{\alpha>0\mid \sum_{m=1}^\infty
\frac{1}{m^{1+\alpha}} h_m <\infty\}\nonumber\\
&=&\sup\{\alpha>0\mid \sum_{m=1}^\infty \frac{1}{m^{1+\alpha}}
h_m =\infty\},
\end{eqnarray}
where we use the (unusual) convention that $\inf\emptyset=\infty$ and $\sup\emptyset=0$.
One could alternatively decide to take the fluctuations into account, and
introduce upper and lower growth exponents in the obvious way [Gu]:
\begin{equation}
\alpha_+(h)=\limsup_{m\to\infty}\frac{\log h_m}{\log m},\  
\alpha_-(h)=\liminf_{m\to\infty}\frac{\log h_m}{\log m}.
\end{equation}
Note that, even if $\alpha_+(h)=\alpha_-(h)$, (1.1) does not necessarily hold.
On the other hand, 
$$
\alpha_-(h)\leq \alpha_0(h)\leq \alpha_+(h).
$$
In addition, if (1.1) holds,
$\alpha(h)=\alpha_+(h)=\alpha_-(h)=\alpha_0(h)$.

Let us write 
$$
\overline{<\psi, U_V^{-m}H_0U_V^m\psi>}=\frac{1}{m}\sum_{k=1}^{m}
<\psi, U_V^{-k}H_0U_V^k\psi>.
$$
The problem is to get upper and lower bounds on this quantity. Upper
bounds of the type  
\begin{equation}
\overline{<\psi, U_V^{-m}H_0U_V^m\psi>}\leq C m^\alpha
\end{equation}
were obtained in [N] under the assumption that $V(t)$ is smooth and the
spectrum of $H_0$ satisfies a gap condition at infinity (see also [J]).  The
proofs are based on adiabatic techniques that do not allow for non-smooth time
dependence and therefore do not apply to kicked systems.  In addition, they
do not provide lower bounds.  One might hope to obtain lower bounds from an
abstract approach initiated in [Gu] and perfected in [C] [L] [BCM]. 
Writing $$ U_V=\int \exp i\lambda\  dE_\lambda
$$
for the spectral decomposition of $U_V$, the main result of this theory (Theorem
6.1 in [L], Theorem 3 in [BCM]) states the following. If the spectral measure
$d<\psi, E_\lambda\psi>\equiv d\mu_\psi(\lambda)$ has Hausdorff dimension 
$\beta$ for some $0\leq\beta\leq1$ then there exists for all $\epsilon>0$ a 
constant $C_\epsilon$, so that, for all
$m\in\N^*$,
 \begin{equation}
\overline{<\psi, U_V^{-m}H_0U_V^m\psi>} \geq C_\epsilon
m^{\gamma_{H_0}(\beta-\epsilon)}. 
\end{equation}
Here  the constant $\gamma_{H_0}$
depends on the spectral properties of $H_0$ in an explicit way [Gu].  The trouble
with such lower bounds is that the information on the spectral measure is very
hard to check:  we do not know of any models where this has been done (other
than numerically or for the trivial case where $\beta=0$ or $1$).  As a result,
we are not aware of any models where lower bounds of the type (1.5) (i.e.
for all $m$) have been proven to hold for $U_V$ with singular
spectrum (other than numerically: see [Gu] for references). We will give such
a model in sections 4-5.  

Actually, the only lower bound we know off is a result of Last ([L], Theorem 7.2)
who shows
that in the Almost Mathieu equation there exist Liouville frequencies so 
that the quantity  $\overline{<\psi_t, X^2\psi_t>}$ (with $\psi_0 =\delta_0$)
has an
upper growth exponent $\alpha_+=2$.
We will strengthen this result in the
following sense.  We will show in sections 3 that, not only in the Almost
Mathieu equation (Theorem 3.2), but more generally in Schr\"odinger operators with
quasi-periodic potentials as well as in the kicked rotor the relevant
dynamical quantity has {\em generically} a {\em mean} growth exponent 
$\alpha_0=2$.  The proof turns out to be very simple and is based on standard
Baire theoretical arguments, sometimes irreverentially refered to as
generic nonsense (section 2). It seems therefore that the occurence of ballistic peaks
close to resonance is a rather common phenomenon.   

The growth exponents $\alpha_+$ and $\alpha_0$ only give lower bounds along a 
subsequence of times.  To display a model in which non-trivial lower bounds 
{\em at all times}
can be obtained, we present in section 5 a tight-binding Hamiltonian with 
electric field $E(t) = E_0 + E_1 \cos\omega t$ and show that, for suitable
choices of $\omega, E_1> E_0$ and provided the off-diagonal matrix elements
satisfy $a_n\sim |n|^{-\nu} (\nu>3/2)$, one has for all $\ell$ and $m$
$$
\overline{<\delta_\ell, U_T^{-m}N^2U_T^m\delta_\ell>}\geq C_\epsilon
m^{\frac{2}{\nu+\frac{1}{2}}-\epsilon},
$$
where $U_T$ is the Floquet operator of the theory (Theorem 5.1).
This result is obtained by 
remarking that the model is equivalent to a kicked
linear rotor and upon using results of [DBF].  A detailed analysis of the 
spectral properties of $U_T$ is also given.

\sect{Two generic results}
In many situations of interest either the Hamiltonian $H(\omega)$ or the
Floquet operator $U(\omega)$ depend on one or several real variables in such a
way that for a dense set of values of $\omega$ a ``resonance" occurs. We will
give examples in the next sections.  For these values the spectrum is
absolutely continuous and much is then known about the dynamical behaviour of
the system.  We show in this section two abstract results permitting to use this
information to draw conclusions on the nature of the spectrum and on the
asymptotic dynamical behaviour of the system for generic values of $\omega$,
``off resonance". 

The term generic is used here in the topological sense. It means for $\omega$
belonging to a dense $G_\delta$-set.  We recall that a $G_\delta$ set 
is a countable
intersection of open sets and that, on $\R^n$, an intersection of two dense
$G_\delta$-sets is still a dense $G_\delta$-set. In addition, dense
$G_\delta$-sets are locally uncountable. It should nevertheless be stressed
that dense $G_\delta$-sets can very well be of zero measure; in fact, in many
applications where the set can be described explicitly, this is the case.

Let ${\cal H}$ be a Hilbert space 
and $A$ a
self-adjoint operator on ${\cal H}$ with domain  ${\cal D}(A)$. Let $H(\omega)$
be a family of self-adjoint operators on ${\cal H}$ with common domain $\cal
D$: we will assume throughout it is continuous in the strong resolvent sense. 
We will write $U_t(\omega)=\exp -itH(\omega)$ for the corresponding unitary
one-parameter group, which is then strongly continuous, uniformly in $t$ on
compacta [RSI].  

\begin{prop} Let ${\cal C} =\{\omega\mid
\sigma_{pp}(H(\omega)) =\emptyset\}$.  Then ${\cal C}$ is a $G_\delta$ set.
In particular, if ${\cal C}$ is dense, it is a dense $G_\delta$.   
\end{prop}

In the following, we suppose
\begin{equation}
H(\omega) = H_0 + V_\omega,
\end{equation}
where $H_0$ is self-adjoint with domain ${\cal D}$ and the $V_\omega$ form a
strongly continuous family of bounded self-adjoint operators.  

\begin{prop} Suppose ${\cal D}(A)\cap{\cal D}$ is dense and

(i) $\forall\omega\in\R^n, \forall t\in \R\,\  U_t(\omega){\cal D}(A)\subset
{\cal D}(A)$;

(ii) $\forall \omega\in\R^n,\ [H(\omega), A]$ is relatively $H_0$ bounded and
the map $\omega\in\R^n\to [H(\omega), A](H_0+i)^{-1}\in{\cal L(H)}$ is strongly
continuous.

Suppose in addition that for some $\psi\in{\cal D}(A)\cap{\cal D}$ there exists
a $\beta_0>0$ so that for all $\omega$ in a dense subset ${\cal R}_\psi$
of $\R^n$ 
\begin{equation}
\beta_0=\sup\{\alpha >0\mid \int_{1}^\infty \frac{dt}{t^{1+\alpha}} 
\overline{<U_t(\omega)\psi, A^*AU_t(\omega)\psi>}=\infty\}. 
\end{equation}
  Then for all $\omega$ in a dense $G_\delta$ set ${\cal S}_\psi$ one has:
\begin{equation}
\beta_0\leq\sup\{\alpha >0\mid \int_{1}^\infty \frac{dt}{t^{1+\alpha}} 
\overline{<U_t(\omega)\psi, A^*AU_t(\omega)\psi>}=\infty\}. 
\end{equation}  
\end{prop}
{\bf Remark:} (i) Proposition 2.1 is Theorem 1.1 in [Si]. We have preferred to
give an
independent proof based on the use of the Wiener theorem, since the same type
of argument serves  to show Propositions 2.2-2.4; (ii) As in [Si], $\R^n$ can 
be replaced by a complete metric space;
(iii)
The time-average appearing in (2.2) and (2.3) can be omitted for the abstract
argument, but often appears in applications.
\vglue 0.2cm
\noindent{\bf Proof of Proposition 2.1:} 
Let $\psi_i, i\in\N$ be an orthonormal basis of ${\cal H}$. Consider for each
$i\in\N,\ T\in [1,\infty[$ the continuous functions $$
R_T^i:\omega\in\R^n\to [\frac{1}{T}\int_0^T\mid<\psi_i,
U_t(\omega)\psi_i>\mid^2 dt]\in \R.
$$
Then Wiener's theorem immediately implies that
$$
{\cal C} = \bigcap_{i,n\in\N}\bigcup_{T\in\N} \{\omega\mid R_T^i(\omega)<{1\over n}\}.
$$
This proves the proposition.\endproof

\vglue 0.2cm
\noindent{\bf Proof of Proposition 2.2:}  We first show 
that the maps 
\begin{equation}
\omega\in\R^n\to<U_t(\omega)\psi,A^*A U_t(\omega)\psi>\in\R
\end{equation}
are continuous for all $\psi\in{\cal D}(A)$.  For that purpose, compute
for $\psi\in{\cal D}(A)\cup {\cal D}$
$$
(AU_t(\omega) -AU_t(\omega'))\psi=[A, U_t(\omega)]\psi - [A, U_t(\omega')]\psi
+(U_t(\omega)-U_t(\omega')A\psi,
$$
and
$$
[A,U_t(\omega)] = U_t(\omega)i\int_0^t ds\ 
U_{-s}(\omega)[H(\omega),A]U_s(\omega).
$$
Continuity of (2.4) now follows easily. For all $\epsilon>0$ and for all
$\omega\in{\cal R}_\psi$, (2.2) implies 
\begin{equation}
\sup_T \int_1^T\frac{dt}{t^{1+\beta_0-\epsilon}}
\overline{<U_t(\omega)\psi, A^*AU_t(\omega)\psi>}=\infty.
\end{equation}
It is a standard corollary of the Baire cathegory theorem ([S], Corollaire 4, p
325) that (2.5) continues to hold for all $\omega$ in
 a dense $G_\delta$ set ${\cal S}_\psi^\epsilon$ containing ${\cal R}_\psi$ and
hence  
\begin{equation}
\int_1^\infty\frac{dt}{t^{1+\beta_0-\epsilon}}
\overline{<U_t(\omega)\psi, A^*AU_t(\omega)\psi>}=\infty, \ \forall \omega\in
{\cal S}_\psi^\epsilon.
\end{equation}
Define ${\cal S}_\psi=\cap_{n\in\N^*}{\cal S}^{1/n}_\psi$.  Then (2.6) implies
that for all $\omega\in {\cal S}_\psi$ (2.3) holds. \endproof

Analogous results hold when dealing with iterates of a unitary operator
$U(\omega)$. We state the results without proofs, which are completely
analogous.  

\begin{prop} Let $\omega\in\R^n\to U(\omega)$ be a strongly continuous map.  If
the pure point spectrum of $U(\omega)$ is empty on a dense set of $\omega$, then
this remains true on a dense $G_\delta$ set.
\end{prop}

\begin{prop} Suppose

(i) $\omega\in\R^n\to U(\omega)\in {\cal U(H)}$ is strongly continuous;

(ii) $\forall\omega\in\R,\ U(\omega){\cal D}(A)\subset {\cal D}(A)$;

(iii) $\forall \omega\in\R^n,\ [U(\omega), A]$ is bounded and the map
$\omega\in\R^n\to [U(\omega), A]\in{\cal L(H)}$ is strongly continuous.

Suppose in addition that for some $\psi\in{\cal D}(A)$ there exists a
$\beta_0>0$ so that
\begin{equation}
\beta_0=\sup\{\alpha>0\mid \sum_{m=1}^\infty\frac{1}{m^{1+\alpha}}
\overline{<U(\omega)^m\psi, A^*AU(\omega)^m\psi>}=\infty\},
\end{equation}
for all $\omega$ in a dense subset ${\cal R}_\psi$ of $\R^n$.  Then for all
$\omega$ in a dense $G_\delta$ set ${\cal S}_\psi$
\begin{equation}
 \beta_0\leq \sup\{\alpha>0\mid \sum_{m=1}^\infty\frac{1}{m^{1+\alpha}}
\overline{<U(\omega)^m\psi, A^*AU(\omega)^m\psi>}=\infty\} 
\end{equation}
 \end{prop}

\section{Applications}
\subsection{The kicked rotor}
It is hard to get rigorous results on the dynamics or on the
dynamical or spectral stability for the kicked rotor [IS].
For the quantum model, with Floquet operator
$$
U_V(\omega) = \exp -i\omega \frac{P^2}{2} \exp-iV(X)
$$
acting on $L^2(\To)$ the only rigorous result we know of is due to Casati-Guarneri
[CGu].  In the following theorem we give both an improvement on this result and
a simplified proof of it.  

\begin{th} Suppose $V\in L^2(\To),\ V'\in L^\infty(\To)$ and that
$\forall \pi\omega\in \Q$, the spectrum of $U_V(\omega)$ is purely
absolutely continuous.  Then there exists a dense $G_\delta$ subset ${\cal S}$
of $\R$ so that 

(i) $\forall \omega \in {\cal S},\ \sigma_{pp}(U_V(\omega))=\emptyset$;

(ii) $\forall \omega\in {\cal S}$, for all momentum eigenstates
$\psi_\ell(x)=\exp i2\pi\ell x,\ \ell\in\Z,$ 
\begin{equation}
\sup\{\alpha>0\mid
\sum_{m=1}^\infty\frac{1}{m^{1+\alpha}}\overline{<\psi_\ell,
U_V(\omega)^{-m} P^2 U_V(\omega)^m\psi_\ell>}=\infty\}=2, 
\end{equation} 
i.e. the mean growth exponent of the time-averaged	 kinetic energy is
$2$. \end{th}

The proof of this result is given below. Let us first point out that the hypothesis
on $\sigma(U_V(\omega))$ for $\pi\omega\in\Q$ is proven to hold for a generic set of
$V$ in [CG].  Note
that part (i) of Theorem 3.1 is Theorem 2 of [CGu]. Actually, the statement of Theorem 2 
in
[CGu] is weaker than the one given here, but Theorem 3.1(i) immediately
follows from their proof.   Part (ii) shows that for $\omega$ belonging to
$\cal S$, the time-averaged kinetic  energy displays ``almost ballistic peaks"
at sufficiently many  time scales to force the mean growth exponent  $\alpha_0$
to equal $2$.  We have no precise information on
how  frequent  these time scales are, but expect them to be extremely
rare, and hence difficult to  detect numerically. 
Since the proof of Theorem 3.1 is based on the abstract nonsense of the
previous section, we have no handle on the set $\cal S$ either. It should nevertheless
be thought of as a set of Liouville numbers.  The main open problem in the
kicked rotor is to show that for $\omega$ a quadratic irrational and for
sufficiently high coupling constant dynamical localization occurs: we
unfortunately have nothing to say on that.  
 
\noindent{\bf Proof of Theorem 3.1:}  It will be sufficient to check 
$$U_V(\omega)=
\exp-i\omega \frac{P^2}{2} \exp-iV(X)
$$ 
satisfies the conditions of Propositions 2.3 and
2.4 with $A=P$. That $\omega\to U_V(\omega)$ is strongly continuous is
immediate and since
$$
PU_V(\omega)\psi=U_V(\omega)P\psi - U_V(\omega)V'\psi,
$$
hypotheses (ii)-(iii) of Proposition 2.4 follow as well.  Hence part (i) of
Theorem 3.1 follows from Proposition 2.3.  As for part (ii), since for all
$\pi\omega\in\Q, \ \sigma_s(U_V(\omega))=\emptyset$, the results of
Last and Guarneri [G][L] show that $\forall \pi\omega\in \Q$ and $\forall
\psi_\ell \ \exists C_\ell(\omega)$ so that  
\begin{equation}
\overline{<U(\omega)^m\psi_\ell, P^2
U(\omega)^m\psi_\ell>}\geq C_\ell(\omega) m^2.
\end{equation}
For all $\pi\omega\in\Q$, the mean growth exponent is therefore equal to 
$2$. The result then follows from Proposition 2.4, provided we show that, 
for all $\omega$, the mean growth exponent is smaller or equal than two.
This is obvious upon remarking that
$$
U_V^{-m}P^2U_V^m = \bigl[P-\sum_{k= 0}^{k=m-1} U_V^{-k}V'U_V^k\bigr]^2.
$$
\endproof

Note that there are other models, such as the kicked Harper model, where
resonances occur, and the results can be adapted to such cases as well.

\subsection{Quasiperiodic potentials}

Let $\Omega$ be an $n+1$ by $n$ matrix and $W$ a continuous
$\Z^{n+1}$-periodic real-valued function on $\R^{n+1}$. Define, for $x\in\R^n$,
$$
V_\Omega(x) = W(\Omega x).
$$
Note that, if $\Omega$ has only rational entries, there is a sublattice of 
$\Z^n$ over which $V_\Omega$ is periodic.  As a result, the Schr\"odinger
operator
$$
H_\Omega = -\Delta + V_\Omega
$$
on ${\cal D}(H_\Omega) = {\cal D}(-\Delta)$ has purely absolutely continuous
spectrum. For $\psi\in{\cal D}(X)$ one then has that $<\psi_t, X^2\psi_t>^2
\sim t^2$ (see [AK] for a detailed proof of this folk theorem).  Identifying
$\Omega$ with an element of $\R^{n(n+1)}$, it is easy to see that the
hypotheses of Proposition 2.1 and Proposition 2.2 are satisfied by $H_\Omega$,
taking $A=X$ and $\psi\in{\cal D}(-\Delta)\cap {\cal D}(X)$.  As a result,
there is a $G_\delta$ dense set of $\omega$ so that the mean growth exponent of
$<\psi_t, X^2\psi_t>$ is $2$ and, in particular, for all strictly
monotonic functions $F(T), F(T)\to\infty,\  \exists C>0, T_k$,  so that
\begin{equation} 
<\psi, {\rm e}^{iH_\Omega T_k}X^2{\rm e}^{-iH_\Omega T_k}\psi>  
\geq C \frac{T_k^2}{F(T_k)}.
\end{equation}
This is of course easily adapted to discrete Schr\"odinger operators and
holds for the time-averaged quantities as well.  In fact, combining the above
with Theorem 7.1 in [L], one gets the following result on the almost Mathieu
equation.   
\begin{th} Let $H_{\lambda,\theta}(\omega) = -\Delta + 
\lambda \cos 2\pi(\omega n + \theta)$ on $\ell^2(\Z)$, with $\mid\lambda\mid>2$
and $\theta \in\R$.  Then there exists a dense $G_\delta$ set of $\omega$ 
so that

(i) All spectral measures of $H_{\lambda, \theta}(\omega)$ are zero Hausdorff
dimensional;

(ii) For all $\psi$ in a dense subset of $\ell^2(\Z)$ contained in 
${\cal D}(X)$, (3.11) holds and the mean growth exponent of
$<\psi_t, X^2\psi_t>$ equals $2$.
\end{th}

Here $(i)$ is exactly Theorem 7.1 of [L], while $(ii)$ improves 
Theorem 7.2 of [L]. It follows immediately upon taking intersections
of dense $G_\delta$-sets.

\sect{The kicked linear rotor}
\subsection{The model}
Consider a particle moving on a circle $\To$, having therefore the
cylinder $M=T^*\To=\To\times\R$ as phase space.  The Hamiltonian of the
kicked linear rotor [G][Be][B][Ho][O] is  
\begin{equation}
H(x,p,t) = \omega p +\sum_{n\in\Z} \delta (t-n) V(x),
\end{equation}
where $\omega\in\R$ is a fixed rotation number. 
The corresponding Floquet map is  
\begin{equation}
\Phi_{V}(x_0, p_0) = (x_1=x_0+\omega, p_1=p_0 +f(x_0)),
\end{equation}
where $f=-V'$.
Standard canonical quantization of the Hamiltonian in (4.1) leads to
the following Floquet operator on $L^2(\To, dx)$:
\begin{equation}
U_{V}(\omega) =   \exp-i\omega P\exp -iV(X).\\
\end{equation} 

 To have some explicit examples in mind, we define, for
$k\in\N\setminus\{0\}$,  $$ Q_k(x)= {1\over (2\pi i)^k} \sum_{n\in\Z^*} {1\over
n^k} \exp i2\pi n x. $$
Then
$
Q_1(x)= {1\over 2} - x, \quad 0< x<1,
$ and for all $k>1$, $Q_k$ is the unique $k^{{\rm th}}$-order polynomial
satisfying $Q'_k = Q_{k-1},\ \int_0^1 Q_k(x) dx = 0$. Note that 
$Q_k^{(i)}(0)=Q_k^{(i)}(1)$ for all $0\leq i\leq k-2$. Hence 
$Q_k\in C^{(k-2)}(\To)$.  For example,
$
Q_2(x) = -{1\over 2} x^2 + {1\over 2} x - {1\over 12}
$
and 
$
Q_3(x) = \frac{1}{12}x(x-1)(-2x+1)
$.

In order to study the asymptotic behaviour of the momentum variable in
the kicked linear rotor, first note that iterating (4.2) yields 
\begin{equation}
\Phi_{V}^m(x_0,p_0) = (x_m = x_0 + m\omega, p_m= p_0 + S(m,\omega)f(x_0)),
\end{equation}
where
\begin{equation}
S(m,\omega)f(x_0) = \sum_{j=0}^{m-1} f(x_0 + j\omega).
\end{equation} 
Let us point out some immediate and simple features of this model. 

First, if 
$\omega\in\Q$, then typically
$
p_m \sim m
$
as is easily seen from (4.4)-(4.5) and the observation that the motion is then
periodic in the $x$-variable. Indeed, if $\omega=r/s$, with $r, s$ relatively
prime integers, then, for all $\ell\in\N$
$$
\Phi_V^{\ell s}(x_0, p_0) = (x_{\ell s} =x_0, p_{\ell s} = p_0 + 
\ell S(s,\frac{r}{s})f(x_0)).
$$
So $p_{\ell s}\sim \ell$ iff $S(s, \frac{r}{s})f(x_0)\not= 0$.

On the other hand,
if $\omega\in\R \setminus\Q$, then the motion is ergodic in the $x$-variable and
hence one immediately obtains that
$
\lim_{m\to\infty} \frac{1}{m}S(m,\omega)f(x_0) = 0
$
so that
\begin{equation}
\lim_{m\to\infty} \frac{p_m}{m} = 0.
\end{equation}
In short, when $\omega$ is rational, the motion is ballistic on the cylinder,
whereas it is subballistic if $\omega$ is irrational.  

Note however that, even
though $p_m= o(m)$ it is still conceivable that \break
$\sup_m\mid p_m\mid=\infty$.  
Moreover it is well known from the Denjoy-Koksma inequality
that $\liminf_m\mid p_m\mid<C<\infty$ if $f=-V'\in C^1(\To)$.  Hence the
classical dynamics would in such a case correspond to the point $(x_m, p_m)$ 
wandering all over
the cylinder, up and down the $p$ axis, leaving every bounded set at some times
and returning close to the origin at some other times.  This behaviour
is thought of as unusual since it can not occur in systems with time-independent
Hamiltonians [RS III]; one expects it to be reflected in the quantum model
through the presence of singular continuous spectrum, a fact we shall prove
below.  

The asymptotics for the classical dynamics of the kicked linear rotor was
studied in detail in [DBF].  We shall show here that the analysis carries over
to the  quantum model and study the spectral and dynamical (in)stability of the
model.

   Before turning to this task, we show the quantum equivalent of (4.6). We need
a preliminary technical lemma. Let ${\cal D}(P)$ denote the domain of $P$, i.e. 
$$ {\cal D}(P)=\{\psi = \sum c_n \exp i2\pi nx\in L^2(\To,
dx) ; \sum n^2\mid c_n \mid^2<\infty\}.
$$

\begin{lem}If $\psi \in {\cal D} (P)$ and if $V, V'\in L^2(\To)$, then $U_V^m
\psi\in {\cal D} (P)$ $\forall
 m \in \N$. Moreover, 
\begin{equation}
<\psi, U_V^{-m} P U_V^m \psi > = <\psi, S(m,\omega)f(X)\psi> + <\psi, P\psi > 
\end{equation}
and
\begin{equation}
\parallel PU_V^m\psi\parallel^2 = <U_V^m\psi, P^2 U_V^m\psi> = 
\parallel P\exp-iS(m,\omega)V(X)\psi\parallel^2,
 \end{equation} 
where $\parallel\cdot\parallel$ denotes the $L^2$-norm.
\end{lem}
{\bf Proof} : First we compute
\begin{eqnarray}
U_V^m  &=& \exp {- i\omega Pm}\exp{- i \sum^{m-1}_{k = 0} V
(X+k\omega)}\nonumber\\ 
& =&
\exp{- i\sum^m_{k = 1} V (X-k\omega)} \exp {- i\omega Pm}. 
\end{eqnarray}
Now, if $\psi\in {\cal D}(P)$, then 
$\psi \in L^\infty (\To)$, so
$f(x-k\omega) \psi (x-m\omega) \in L^2(\To, dx)$ since $f \in L^2(\To, dx)$. In
conclusion 
$$
P(U_V^m \psi) (x) = \sum^m_{k= 1} f(x-k\omega) (U_V^m \psi) (x) + (U_V^m(P
\psi)) (x) 
$$
belongs to $L^2(\To, dx)$; (4.7) follows immediately from this. Using the
first relation in (4.9), (4.8) follows as well. \endproof

\begin{prop} : If $\psi \in {\cal D} (P)$, if $V, V'\in L^2(\To)$, and if
$\omega$ is irrational one
has, as $m\to\infty$,  
\begin{equation}
{1 \over m} <\psi, U_V^{-m} P U_V^m \psi > \rightarrow 0
\end{equation}
and 
\begin{equation}
{1 \over m^2} \parallel PU_V^m \psi \parallel ^2 \rightarrow 0. 
\end{equation}
\end{prop}
{\bf Proof} : Since 
$\int_{\To} f(x) dx = 0$
 and $f \in L^2(\To, dx)$
 it follows from the Von Neumann mean ergodic theorem that
${\displaystyle
m^{-2} \parallel S(m,\omega)f \parallel ^2 \ \stackrel {m \rightarrow
\infty}{\longrightarrow} 0.}
 $
Hence, using (4.8), one has (4.11)
since $\psi \in L^\infty (\To)$.
The proof of the first statement is similar.\endproof

Clearly (4.10)-(4.11) are the quantum analogs of (4.6). For rational $\omega$,
one sees from (4.7) that, as in the classical case, 
$<\psi,U_V^{-m}PU_V^m\psi>\sim m$ for {\it typical potentials} $V$; 
 we shall say that a potential $V$  is a {\it typical
potential} provided the function $S(s,\frac{r}{s})V$ is not constant on a set of
positive measure for any $r, s$ as above. Note that it follows from our
discussion above that this is equivalent to requiring that $p_m\sim m$
for all rational $\omega$ and for almost all initial values $x_0$. Putting an
appropriate topology on the set of all potentials one can show that the typical
potentials form a dense $G_\delta$ set, but we shall not be interested in that
any further.  It is perhaps more interesting to remark that the potentials $Q_k$
introduced above are typical.  Indeed, it is easy to compute
$$
S(s, \frac{r}{s})Q_k(x) = s^{1-k} Q_k(sx),
$$
from which the conclusion follows. 

\subsection{Spectral (in)stability}    

We are now ready to settle the spectral stability question for the kicked
linear rotor.  The result is summarized in the following theorem. 
\begin{th}  Let $U_V(\omega)=\exp-i\omega P\exp-iV(X)$ with $V, V'\in
L^2(\To)$, i.e. $V\in H^1(\To)$.  Then

(i) For typical $V$ and for all $\omega\in\Q,\
\sigma_{pp}(U_V(\omega))=\emptyset=\sigma_{sc}(U_V(\omega))$; i.e. $L^2(\To) =
{\cal H}_{ac}(U_V(\omega))$;

(ii) If $\omega\in\R\setminus\Q$, then $\sigma_{ac}(U_V(\omega))=\emptyset$ and
in addition either $L^2(\To) = {\cal H}_{sc}(U_V(\omega))$ or 
$L^2(\To) = {\cal H}_{pp}(U_V(\omega))$;

(iii) For typical $V$, there exists a dense $G_\delta$ set of
$\omega$ so that 
$L^2(\To) = {\cal H}_{sc}(U_V(\omega))$  (i.e.
$\sigma_{ac}(U_V(\omega))=\emptyset= \sigma_{pp}(U_V(\omega))$;

(iv) If $V\in H^s(\To)$, $s>1$,   then $L^2(\To) = {\cal H}_{pp}(U_V(\omega))$
(i.e. $\sigma_{ac}(U_V(\omega))=\emptyset= \sigma_{sc}(U_V(\omega))$)
for Lebesgue almost all $\omega$.
\end{th}

\noindent{\bf Proof:} Part (i) is the spectral pendant of the observation
made above that for rational $\omega$, and for typical $V$, the
motion is ballistic: $p_m\sim m$.  If $\omega=r/s$ with $r$ and $s$
relatively prime integers, then it is easy to see that the spectrum of
$U_V(\omega)$ is organized in bands, and has no singular continuous part
[IS][Be].  To see that, for typical $V$, it has no eigenvalues either, we
proceed as follows. First, (4.9) implies  $$
U_V(\omega)^s = \exp -i\sum_{k=0}^{s-1} V(X+ k\frac{r}{s}).
$$
Consequently, it is clear that, provided the function $S(s,\frac{r}{s})V$ is
not constant on a set of positive measure, the spectrum of $U_V(\omega)^s$
is purely absolutely continuous. In that case, the same holds for the
spectrum of $U_V(\omega)$ itself. 

 More interesting is the
case where $\omega\in\R\setminus\Q$.  To prove (ii), recall that the
aforementioned result of Guarneri and Last, and more precisely Theorem 6.1 in
[L] implies that, if the spectral measure of $\psi$
contains an absolutely continuous component, then necessarily  
$$
\overline{<\psi,U_V^{-m}P^2U_V^m\psi>} \geq C m^2,\quad \forall m\in \Z, 
$$ for
some $C>0$.  This being incompatible with (4.11), we conclude that
$\sigma_{ac}(U_V(\omega))$ is empty, proving the first statement in (ii); (iii)
now follows immediately from this, Proposition 4.3, and (i).

The second part of (ii) is easily proven as follows. Suppose
$\lambda\in\sigma_{pp}(U_V(\omega))$ and let $\psi$ be the corresponding
eigenfunction, then 
\begin{equation}
U_V\psi(x)=\exp-iV(x-\omega) \psi(x-\omega)=\lambda \psi(x).
\end{equation}
Taking absolute values on both sides yields 
$$
\mid\psi\mid(x-\omega)=\mid\psi\mid(x),\qquad\forall x\in[0,1[.
$$
Since $\omega$ is irrational, this implies $\mid\psi\mid(x)\equiv{\rm cst}$, so
that $\psi(x)=\exp-iW(x)$ for some $W$.  Reinserting this into (4.12), one has
$$
\exp -iV(x-\omega) = \lambda \exp -iW(x)\exp iW(x-\omega)
$$
and hence
\begin{equation}
U_V(\omega)=\lambda \exp -iW(X)\exp-i\omega P\exp iW(X).
\end{equation}
So $U_V(\omega)$ is unitarily equivalent to $\lambda \exp-i\omega P$, which of
course has pure point spectrum.  In conclusion, if $\omega\in\R\setminus\Q$, we
know that either $\sigma_{c}(U_V(\omega))=\emptyset$, or
$\sigma_{pp}(U_V(\omega))=\emptyset$; the spectrum is therefore either purely
singular continuous or exclusively pure point.  We have already shown that the
first case occurs on a dense $G_\delta$ set.  We now show the second case
occurs as well. 

One expects the spectrum to be pure point for $\omega$ poorly approximated by
the rationals.  To prove this, first remark that it follows from (4.13) that it
suffices to solve
$$
 V(x) = W(x) - W(x-\omega) 
$$
for $W\in L^2(\To)$.  Assuming $V\in H^s(\To), s>1$  and that
$q_{k+1}=O(q_k^{1+\gamma})$ with $1+\gamma<s$ 
such a solution exists as is easily seen upon Fourier transforming
the equation and solving for the Fourier coefficients of $W$. 
It is then easy to see that $U_V(\omega)$ is of the
form (4.13) with $\lambda=1$. Here the $q_k$ are the denominators of the
continuous fraction approximants of $\omega$.  Since the condition $q_{k+1}=
O(q_k^{1+\gamma})$ holds Lebesgue a.e. (Theorem 32 in [K]), (iv) follows. 
\endproof

\noindent{\bf Remark:} A version of part (i) of this result is proven in
[Ho][Be]. 
Absolute continuity
 of the spectrum  implies [Gu][L] that $\forall
\psi\in{\cal D}(P)$, 
$$ \overline{<\psi,U_V^{-m}P^2U_V^m\psi>} \geq C m^2,\quad \forall
m\in \Z, $$ 
a fact we will prove more directly below.  Part (ii) is an
improvement over [Be], where the result is proven on the one hand for all
irrational $\omega$ under the hypothesis that $V\in C^1,\ \int_{\To}\mid
V'\mid(x)dx<2\pi$ and on the other hand, for almost all
$\omega\in\R\setminus\Q$, provided $V\in C^2(\To)$. Our result here shows that
$V'\in L^2(\To)$ suffices and that the condition on the total variation of $V$
can be removed.  Note that some smoothness of $V$ is at any rate needed: it is
mentioned in [Be] and easily confirmed 
that for $V(x)= 2\pi x,\ x\in]0,1[$, the spectrum of
$U_V$ is a.c. even for irrational $\omega$. Note however that, viewed
as a function on $\To$, this potential does not have an $L^2$ derivative
because of the jump discontinuity.  For typical $V\in C^2(\To),\ \int \mid
V'\mid dx < 2\pi$, (iii) is proven in [O] using the results of [CG] and [Be].

\subsection{Dynamical (in)stability}
We are interested in the behaviour of $\overline{<U_V^m\psi,P^2U_V^m\psi>}$.
First remark that the results of section 2 immediately imply  that, 
generically in $\omega$,
the mean growth exponent of this quantity equals $2$.
To get sharper estimates, we now
use the results of [DBF].  
If the Fourier coefficients $v_n$ of $V$ satisfy $v_n\sim|n|^{-(\nu+1)}$
for some $\nu>1/2$, it is an immediate consequence of Theorem 1.1(i)  of [DBF] 
that for a
class ${\cal R}_\infty$ of explicitly described  Liouville $\omega$ and for all
$\epsilon>0$, there exists a constant $C_\epsilon$ so that 
\begin{equation}
\overline{<\psi_\ell,U_V(\omega)^{-m}P^2U_V(\omega)^m\psi_\ell>}\geq
C_\epsilon\ m^{\frac{2}{1+\nu}-\epsilon},\ \forall m,
 \end{equation}
where $\psi_\ell(x) =\exp i2\pi\ell x$.
Here $\omega\in {\cal R}_\infty$ if and only if the denominators $q_k$ of its
continued fraction expansion satisfy 
$$
\forall \gamma>0, \exists R_\gamma>0 \mbox{ so that } q_{k+1}>{{q_k^{1+\gamma}}\over
{R_\gamma}}.
$$
Moreover,
Theorem 1.2(ii) in [DBF] implies that, for the same $\omega$, 
\begin{equation}
\frac{2}{1+\nu} \leq
\alpha_-=\liminf_{m\to\infty}\frac{
\log
\overline{<\psi_\ell,U_V(\omega)^{-m}P^2U_V(\omega)^m\psi_\ell>}}
{\log m}\leq\frac{2}{1/2+\nu}.
\end{equation}
Using (1.5) this implies that 
$
{\rm dim}_H \mu_{\psi_\ell} \leq \frac{1}{1/2+\nu}.
$
 At the same time, and still for the same $\omega$, Theorem 1.1 in [DBF] yields
\begin{equation}
\alpha_+=\limsup_{m\to\infty}\frac{
\log
\overline{<\psi_\ell,U_V(\omega)^{-m}P^2U_V(\omega)^m\psi_\ell>}}
{\log m}=2.
\end{equation}
Summarizing the result loosely, we see that 
$\overline{<\psi_\ell,U_V(\omega)^{-m}P^2U_V(\omega)^m\psi_\ell>}$
oscillates quite a bit, between $m^2$ and $m^{\frac{2}{\nu +1/2}}$, 
staying always at least as high as $m^{\frac{2}{\nu +1}}$.

Note again that the upper bound on the Hausdorff dimension of the spectral
measure $\mu_{\psi_\ell}$ does not exclude the existence of almost ballistic
peaks, a phenomenon already observed in [L] and in Theorem 3.2 above.  

If $V\in C^\infty$, then Theorem 1.2 (ii) of [DBF] shows that,
for all $\omega\in\R\setminus\Q$,
$
\alpha_-=0,
$
implying, via (1.5), that the Hausdorff dimension of $\mu_\psi$ vanishes.  So in this
case, the Hausdorff dimension takes on only two values: it is equal to 
$1$ if $\omega$ is
rational and equal to $0$ otherwise.  At the same time, the results of 
section 2 tell
us that for generic values of $\omega$,
 $\alpha_0=2=\alpha_+$.  This shows in an even stronger way
than the example of section 3.2 that no upper bounds on
$\overline{<\psi,U_V(\omega)^{-m}P^2U_V(\omega)^m\psi>}$ can be expected in
terms of information on the fractal dimension of the spectral measures.

These results show that
$\overline{<\psi,U_V(\omega)^{-m}P^2U_V(\omega)^m\psi>}$ tends to fluctuate
enormously.  It is legitimate to speculate that this is a feature common to
the solution of Schr\"odinger equations with propagators having ``unusual"
spectrum.  Note however that detecting such fluctuations numerically might be
an impossible task: the times at which they are proven to occur (see [DBF]) 
behave
like $q_k$, where the $q_k$ are the denominators of the convergents of
the continued fraction expansion of $\omega$, which grow extremely fast for
the $\omega$ considered.  

To end this section, we should point out that the results in [DBF] actually
concern the asymptotic  behaviour of $<\overline{p_m}>^2$,
where 
$$
<{\overline p_m}>^2 = \int_{\To}\mid {\overline p_m(x_0)}\mid^2 dx_0.
$$  
This is clearly determined by the asymptotic behaviour of the
$L^2$-norm of the time-average of $S(m,\omega)f$. It turns out that this 
quantity has the same asymptotic behaviour as
$\overline{<\psi_\ell,U_V(\omega)^{-m}P^2U_V(\omega)^m\psi_\ell>}$,
as we now briefly show. 
First compute, using (4.8)
\begin{eqnarray}
\bigl[\parallel
S(m,\omega)f(X)\psi\parallel&-&\parallel P\psi\parallel\bigr]^2\\ 
&\leq& <U_V^m\psi, P^2 U_V^m\psi> \leq
\bigl[\parallel S(m,\omega)f(X)\psi\parallel+\parallel P\psi\parallel\bigr]^2.
\nonumber\end{eqnarray}
Averaging over $m$ and a short computation then yield
$$
\bigl[\bigl({1\over m}\sum_{k=1}^m \parallel S(k,\omega)f(X)\psi\parallel^2
\bigr)^{1/2} - \parallel P\psi\parallel\bigr]^2
$$
$$
\leq {1\over m}\sum_{k=1}^m <U_V^k\psi, P^2 U_V^k\psi> \leq
\bigl[\bigl({1\over m}\sum_{k=1}^m \parallel S(k,\omega)f(X)\psi\parallel^2
\bigr)^{1/2} + \parallel P\psi\parallel\bigr]^2,
$$
from which it is clear that the asymptotic behaviour of 
$\overline{<U_V^m\psi, P^2 U_V^m\psi>}$ is determined by the one of 
$
\overline{\parallel S(m,\omega)f(X)\psi\parallel^2}.
$ Assuming
that $\mid\psi\mid(x)=1,\quad\forall x$,
which is in particular true when $\psi$ is a momentum eigenstate
$\psi_\ell(x)$, one simply has
$\parallel S(m,\omega)f(X)\psi_\ell\parallel=\parallel S(m,\omega)f(X)\parallel$
so that the averaged kinetic energy is then determined by the time-averaged
$L^2$-norm of $S(m,\omega)f(X)$.  It turns out that this has the same 
asymptotic behaviour as the $L^2$-norm of the time-average of $S(m,\omega)f$.
To see this, note that a simple
computation shows that $$ \parallel {1\over M}\sum_{m=1}^M
S(m,\omega)f(X)\parallel^2= \sum_{n\in\Z}\mid f_n\mid^2 G_M(n\omega), $$
with
$$
G_M(x)=\frac{1}{4\pi\sin^2\pi x}\mid 1-\frac{\sin\pi Mx}{M\sin\pi x}\exp i\pi Mx
\mid^2
$$
whereas
$$
 {1\over M}\sum_{m=1}^M \parallel S(m,\omega)f(X)\parallel^2=
\sum_{n\in\Z}\mid f_n\mid^2 H_M(n\omega),
$$
with
$$
H_M(x)=\frac{1}{2\sin^2\pi x}\bigl[1-\frac{\cos \pi x(M+1)\sin \pi x
M}{M\sin\pi x}\bigr]. $$
It then suffices to notice that the proofs of [DBF] are entirely based on the
following estimates for $G_M(x)$ (Lemma 2.1 in [DBF])
$$
c/\sin^2\pi x\leq G_M(x) \leq C/\sin^2\pi x,\quad\forall 1/M \leq x \leq 1/2
$$
and
$$
c M^2\leq G_M(x)\leq  CM^2\quad \forall 0\leq x\leq 1/M,
$$
with the upper bound holding for all $x\in\R$.  The same inequalities hold
for $H_M(x)$ as well, as is easily checked.   It follows that the quantum and
classical dynamics in this model are essentially the same. This is not too
surprising if one compares $p_m=p_0 + S(m,\omega)f(x_0)$ to (4.7), which can be
rewritten suggestively as (see also [O])
$$
P_m \equiv U_V^{-m} P_0 U_V^m = P_0 + S(m,\omega)f(X_0).
$$

\sect{A one-band tight binding model in a time-dependent electric field}
In this section we turn our attention to the one-dimensional tight-binding
model with time-dependent electric field $E(t)$ given by
\begin{equation}
[H(t)\psi]_n =\sum_{k\in\Z} a_{n-k} \psi_k + E(t)n\psi_n.
\end{equation}
Here the off-diagonal matrix elements $a_n$ of the Hamiltonian are assumed to
belong to $\ell^2(\Z)$ and $a_n=a_{-n}$.  The corresponding time-dependent
Schr\"odinger equation
\begin{equation}
i\partial_t\psi_t = H(t)\psi_t,
\end{equation}
can be solved explicitly in several ways [DK].  We proceed as follows. 
Fourier transforming of (5.2) yields
$$
i\partial_t\hat\psi_t = \hat H(t)\hat\psi_t,
$$
where, for $x\in[0,1[\equiv \To$,
\begin{eqnarray*}
\hat\psi(x) &=&\sum_n\psi_n\exp i2\pi n x \in L^2(\To,dx),\\
\hat a(x) &=&\sum_n a_n\exp i2\pi n x \in L^2(\To,dx),\\
\hat H(t)&=&\frac{E(t)}{2\pi} P + \hat a(x),\ P=\frac{1}{i}\frac{d}{dx}.
\end{eqnarray*}

We chose to use the unusual notation ``$x$" for the quasi-momentum in order
to bring the analogy with the previous section out more clearly, as follows. 
Solving the Schr\" odinger equation yields
\begin{equation}
\hat\psi_t = U_t\hat\psi_0,
\end{equation}
where
\begin{equation}
U_t = \exp -i G(t) P \exp -iW_t(X),
\end{equation}
and
\begin{equation}
G(t) = \frac{1}{2\pi}\int_0^t E(s) ds;\quad W_t(x) = \int_0^t \hat a(x+G(s))
ds.
\end{equation} 
We now concentrate on smooth periodic field amplitudes $E(t)$ of period $T$ for
which we write 
\begin{equation}
E(t+T) = E(t); \ E(t) = E_0 + e(t); \ e(t) = \sum_{n\in\Z^*}e_n\exp
i\frac{2\pi}{T}nt.
\end{equation}
The Floquet operator $U_T$ is then given by
\begin{equation}
U_T=\exp -i \tau P \exp -i W_T (x),\quad \tau = \frac{E_0T}{2\pi}.
\end{equation}
The analogy with the previous section, and in particular with (4.3) is now
completely clear.  We shall be interested in the asymptotic behaviour of 
\begin{equation}
<\psi_t, N^2\psi_t> = \sum_{n\in\Z}n^2\mid\psi_{tn}\mid^2 = \frac{1}{(2\pi)^2}
<\hat\psi_t, P^2\hat\psi_t>,
\end{equation}
for $\psi_0\in {\cal D}(N)$.  For this expression to make sense, we need
that $\psi_t\in{\cal D}(N)$. A
sufficient condition to ensure  that ${\cal D}(N)$ is invariant under $U_t$ is
given by Lemma 4.1:  $na_n\in\ell^2(\Z)$.  

Note first that, if $E_0=0$, it is well known that $<\psi_t, N^2\psi_t>\sim
t^2$, except if $e(t) = E_1\cos\frac{2\pi}{T}t$, 
$a_n = \delta_{1,n} + \delta_{-1,n}$ and
with very special choices of $T$ and $E_1$ [DK].  If, on the other hand,
$e(t)=0$, one readily sees that $\sup_t<\psi_t, N^2\psi_t><C$.  We will
consider the intermediate situation where $E(t) = E_0 +
E_1\cos\frac{2\pi}{T}t$ and $E_1>E_0>0$, to obtain the following result.
\begin{th} Suppose $a_n\geq c\mid n\mid ^{-\nu}$ for some $c>0$ and $\nu>3/2$.
Then there exists a set of Liouville $\omega$ so that the following holds.
Given $\omega$ in this set and given $E_0$, there exists a countable and dense
set of values of $E_1$ in $]E_0,\infty[$ so that for all $\epsilon >0$
and $\forall \ell\in\Z$, there exists $C>0$ so that,   
$$
\overline{<\delta_\ell, U_T^{-m}N^2U_T^m\delta_\ell>}\geq C
m^{\frac{2}{\nu+ \frac{1}{2}}-\epsilon},
$$
Here $C$ depends on $\epsilon, \ell, E_0,\omega$ and $E_1$.  
\end{th}
Note that, as $\nu\to 3/2$, the exponent approaches $1$ from below, so that
under our hypotheses the motion is always subdiffusive.  The lower bound on 
$\nu$ ensures that the evolution leaves the domain
of $N$ invariant.  

\noindent{\bf Proof:} Comparing (5.4) to (4.3), it is clear that, in order
to apply the results of section 4.3, we need to get a lower bound on the 
Fourier coefficients $w_n=\overline{w}_{-n}$ of $W_T(x)$:
$$
w_n = a_n \int_0^T \exp in[E_0t + \frac{E_1}{\omega}\sin\omega t]\ dt.
$$
To control the integral, we use a stationary phase argument.  Since we 
assume $E_1>E_0$, the phase has two stationary points $0<t_-<t_+<T$:
$$
E_0 + E_1\cos\omega t_\pm =0.
$$
Taking small open intervals $\Delta_\pm\subset ]0,T[$ around $t_\pm$
so that $\Delta_+\cap\Delta_-=\emptyset$, one can construct a smooth
partition of unity on $\R/T\Z$ so that
$$
\varphi_- + \varphi_0 + \varphi_+ = 1,
$$
where 
${\rm supp}\varphi_\pm\subset\subset\Delta_\pm,\ 0\leq\varphi_\epsilon\leq1
\in C^\infty$ and where 
$\varphi_{\pm}(t) =1$ on a small subinterval of $\Delta_\pm$ containing
$t_\pm$.  Then
$$
\int_0^T\exp i2\pi nG(t) \ dt = I_0^n + I_+^n +I_-^n,
$$
with obvious notations. Since we know that, on the support of $\varphi_0$,
$\mid G'(t)\mid$ is bounded away from $0$, an integration by parts shows
that $\mid I_0^n\mid = O(|n|^{-1})$.  For $I_{\pm}^n$, a stationary
phase argument shows that
$$
I_{\pm}^n = \sqrt{\frac{1}{n\mid G''(t_{\pm})\mid}}\exp i2\pi nG(t_{\pm})
\exp i\frac{\pi}{4}{\rm sgn}G''(t_{\pm}) + o(\mid\! n\!\mid^{-1/2}).
$$
Since $G''(t_+) = -G''(t_-)>0$ one finds
\begin{eqnarray*}
\int_0^T\exp i2\pi n G(t) \ dt = \sqrt{\frac{2\pi}{n\omega(E_1^2-E_0^2)^{1/2}}}&
\exp i\frac{\pi}{4}\exp i2\pi nG(t_{+})
[1- i \exp in \Phi(E_1)] \\&+ o(|n|^{-1/2})
\end{eqnarray*}
where, for $E_0$ and $\omega$ fixed,
$$
\Phi(E_1) = E_0(t_- - t_+) + \frac{2E_1}{\omega} \sin\omega t_-. 
$$
Since $\Phi$ is a monotonically increasing continuous function of $E_1$, 
mapping $]E_0, \infty[$ onto $]0,\infty[$, one can
choose $E_1$ so that $\Phi(E_1) = 2\pi \frac{p}{q}$, with $p,q\in\N$ and $q$
different from $0$  modulo $4$, which guarantees that
$$
\inf_n \mid [1- i\exp in \Phi(E_1)]\mid >0
$$
from which one concludes that
$$
\mid \int_0^T\exp i2\pi G(t) \ dt\mid \geq C \frac{1}{\sqrt n}.
$$
This implies $w_n\geq \mid n\mid^{-(\nu+1/2)}$, so that (5.8) and (4.14)
yield the result of the Theorem. \endproof

\end{document}